\documentclass{article}

\usepackage{arxiv}
\usepackage{authblk}
\usepackage{graphicx}
\graphicspath{{./figs/}}
\usepackage{color}
\usepackage{array}
\usepackage{multirow}
\usepackage{threeparttable}
\usepackage{booktabs}
\usepackage{stfloats}
\usepackage{amsmath}
\usepackage{csquotes}
\usepackage{subfigure}
\usepackage{fancyhdr} % for header and footer
\usepackage[symbol]{footmisc}

\usepackage{IEEEtrantools}
\usepackage{listings}
\usepackage{tikz}
\usepackage{afterpage}
\usepackage{url}
\usepackage{soul}
\usetikzlibrary{automata,arrows,positioning,arrows.meta}
\newcolumntype{C}[1]{>{\centering\let\newline\\\arraybackslash\hspace{0pt}}m{#1}}

\AtBeginDocument{%
	\providecommand\BibTeX{{%
			\normalfont B\kern-0.5em{\scshape i\kern-0.25em b}\kern-0.8em\TeX}}}

\title{On the RTL Implementation of FINN Matrix Vector Compute Unit}

\author[1]{Syed Asad Alam and David Gregg}
\affil[1]{School of Computer Science and Statistics\\
	The University of Dublin, Trinity College\\
	Dublin, Ireland
	\thanks{\texttt{\{syed.asad.alam,david.gregg\}@tcd.ie}}}

\author[2]{Giulio Gambardella and Michaela Blott}
\affil[2]{Xilinx Research Labs\\
	Ireland
	\thanks{\texttt{\{giuliog,mblott\}@xilinx.com}}}

\author[3]{Thomas Preusser}
\affil[3]{Xilinx Research Labs\\
	Germany
	\thanks{\texttt{tpreusse@xilinx.com}}}

\begin{document}
	
	\maketitle
	
	\begin{abstract}
		FPGA-based accelerators are becoming increasingly popular for deep neural network inference due to their ability to scale performance with increasing degree of specialization with dataflow architectures or custom data type precision. In order to reduce the barrier for software engineers and data scientists to adopt FPGAs, C++- and OpenCL-based design entries with high-level synthesis (HLS) have been introduced. They provide higher abstraction  compared to register-transfer level (RTL)-based design. HLS offers faster development time, better maintainability and more flexibility in code exploration,  when evaluating several options for multi-dimension tensors, convolutional layers or different degrees of parallelism. For this reason, HLS has been adopted by DNN accelerator generation frameworks such as FINN and hls4ml.
		
		In this paper, we present an alternative backend library for FINN, leveraging RTL.  We investigate and evaluate, across a spectrum of design dimensions, the pros and cons of an RTL-based implementation versus the original HLS variant. We show that for smaller design parameters, RTL produces significantly smaller circuits as compared to HLS. For larger circuits, however, the look-up table (LUT) count of RTL-based design is slightly higher, up to around $15\%$. On the other hand, HLS consistently requires more flip-flops (FFs) (with an orders-of-magnitude difference for smaller designs) and block RAMs (BRAMs) ($2\times$ more). This also impacts the critical path delay, with RTL producing significantly faster circuits, up to around $80\%$. Furthermore, RTL also benefits from at-least a $10\times$  reduction in synthesis time. Finally the results were practically validated using a real-world use case of a multi-layer perceptron (MLP) network  used in network intrusion detection. Overall, since HLS frameworks code-generate the hardware design, the benefits of the ease in the design entry is less important. As such, the gained benefits in synthesis time together with some design dependent resource benefits, might make the RTL abstraction an attractive alternative.

	\end{abstract}

	
	\section{Introduction}
	\label{sec:intro}
	
	Deep convolutional neural networks, referred to simply as CNNs, have shown a tremendous growth in the past many years with networks now having millions of parameters. E.g., AlexNet, VGGNet and ResNet-152 have $62M$, $138M$ and $60.3M$ parameters, respectively \cite{Krizhevsky2017M, Simonyan2014X, He2015C}. The computational cost of a CNN is primarily derived from \textit{convolution layers}. These convolution operations are typically performed between input matrices, which can be as large as $224\times 224$ with multiple channels as in the popular ImageNet category\cite{Deng_2009C_ImageNet}, and filter kernels, which are typically small on the order of $5\times 5$ \cite{Szegedy_2015C_GoogleNet} or $3\times 3$ \cite{Simonyan2014X}. 
	%Albeit, earlier networks like AlexNet did feature larger filter kernels of size $11\times11$ \cite{Krizhevsky2017M}. 
	The implied convolutional compute amounts to many dot products between the filter kernels of all output channels and correspondingly sized, overlapping tiles of the input across the full depth of input channels. Many networks conclude the convolutions with a small number of fully connected layers. These layers, indeed, compute one large dot product for each output channel but without any stencil tiling. In consequence, convolutional layers dominate the total execution time of modern CNNs. In total, these compute requirements are challenging in CNN deployments especially on resource-constrained devices, say, for enabling Internet-of-Things (IoT) use cases both embedded and at the edge.
	The deployment of such networks on edge devices is critical to enable technologies like smart cities, driver-less cars and other autonomous systems \cite{Satyanarayanan2017J} as it is neither energy-efficient to submit data from an edge device for remote cloud processing nor conducive for the real-time processing required by such state-of-the-art applications.
	
	%%% Efforts to circumvent DNN computational cost
	The challenging demands of the desired CNNs must be mitigated on embedded devices with limited processing and memory capacity. It is important to reduce the number of layers or parameters, and the memory required to store them. There are various ways in which these aspects of a network can be reduced. One of the ways is referred to as pruning. It involves replacing insignificant parameters in weight tensors by zero. Pruning can be performed at different levels of granularity ranging from pruning individual weights to complete channels \cite{Ju_2016X, Persand_2021J}.
	
	%% Quantization
	Another orthogonal measure of reducing the memory footprint of such networks is quantization. It means that the data inside a network (i.e., input and output activations, weights and biases) are represented using shorter word lengths. 
	Admissible word lengths have been explored thoroughly for fixed- and floating-point number systems. Capotondi et al. \cite{Capotondi2020J} propose a CNN inference library, called CMix-NN, for integer-only mixed low-precision quantization of weights and activations at $(8,4,2)$-bits. The library is optimized for the ARM instruction set with vector arithmetic extensions and tested for the deployment on microcontroller targets like the STM32H7. Garofalo et al. \cite{Garofalo2019J} also propose a library of kernels for quantized neural network (QNN) inference, called PULP-NN, using a homogeneous quantization scheme with $(8,4,2,1)$-bits. They optimize for a cluster of low-power RISC-V processors by exploiting $4\times8$-bit SIMD MAC instructions and bitwise extension operations in order to benefit from precisions below 8~bits. Bruschi et al. \cite{Bruschi2020C} extend PULP-NN by targeting the acceleration of mixed-precision DNNs.
	Hubara et al. \cite{Hubara2017J} present the quantization with various low-bit representations as small as $1$~bit per weight and activation to allow for the use of bitwise operations during the forward pass. A further contribution in the field of these \emph{binary neural networks} (BNNs) is reported by Rastegari et al. \cite{Rastegari2016X}. They present two approximations of DNNs by firstly representing all weights with binary values and secondly approximating both weights and inputs to the convolutional and fully connected layers by binary values. Their interpretation of the two values as $\pm1$ motivates their term \emph{XNOR network} as inspired by the resulting elementary multiplication.
	
	%%% Brief Introduction of FINN
	Field-programmable gate arrays (FPGAs) are a very suitable platform to implement fixed-point systems due to the presence of fast multipliers and adders. Their ability to customize datapaths down to the bit level has made them an attractive target and the technology of choice for very-low-precision QNNs and, of course, BNNs. One framework exploiting this technological match is FINN, which is able to build FPGA-based accelerators for QNNs. While initially limited to BNNs \cite{Umuroglu_2017C}, FINN was later extended to non-binary word lengths \cite{Blott2018J}. The prototypes designed using this framework were demonstrated to improve performance measures, such as latency, throughput, Top-1, Top-5 and mean average precision (mAP) accuracies, with respect to the state of the art.
	The FINN framework uses C++-based high-level synthesis (HLS). HLS allows software engineers and scientists, not well versed with hardware description languages (HDLs) like Verilog and VHDL to describe digital systems. They can, thus, target hardware platforms like FPGAs and ASICs using familiar languages like C++ or OpenCL \cite{Meeus_2012J, Martin_2009J, Sarkar_2009J}. Designs described in these high-level languages are translated to traditional HDLs using dedicated compilers. 
	The time required to develop and design a digital system using HLS is considerably less than what is needed for a genuine HDL implementation. The general reduction of code complexity is a benefit for design space exploration, optimization, and maintenance. Also, the debugging of a design can ideally be performed on a higher functional abstraction level, which is more accessible to the application engineer and considerably faster than a simulation on the register-transfer level (RTL).
	Finally, HLS offers more flexibility in code exploration, which is powerful for example when evaluating the many options for ordering different dimensions in tensors, convolutional layers or different degrees of parallelism.
	
	%%% What are we proposing
	On the other side, adding an extra layer of abstraction and translation also incurs costs in terms of design implementation times, hardware resource cost and the predictability of design quality. In this work, we analyse HLS- and RTL-based implementations of the core compute unit of the accelerator designed for the FINN framework to evaluate the implied trade-offs. This core unit performs matrix-vector multiplications using different types of compute engines and uses AXI-Stream-based interfaces for communication. We use Xilinx's Vivado and Vivado\,HLS for RTL and HLS implementations, respectively.
	
	The main contributions of this work are:
	
	\begin{itemize}
		\item RTL implementation of the key FINN compute component, the MVU, available in open source at \linebreak \url{https://github.com/asadalam/FINN_MatrixVector_RTL}.
		\item Systematic measurement and analysis of resource utilization, critical path delay and synthesis time for designs realized using HLS and RTL. We show that:
		\begin{itemize}
			\item HLS uses more FPGA resources for smaller designs and suffers from complex multiplexer structures when processing large input streams using a limited number of compute units.
			\item The LUT usage converges between HLS and RTL as the core compute unit increases in size but HLS consistently uses more flip-flops and block RAMs (BRAMs).
			\item The RTL implementation results in a significantly reduced critical path delay across different parameters and for the three different types of compute units as compared to HLS
			\item HLS takes at least $10\times$ more synthesis time as compared to RTL synthesis and is the limiting factor towards synthesizing and analyzing designs with large numbers of compute units.
		\end{itemize}
		\item Implementation and analysis of a multi-layer perceptron (MLP) network, consisting of 4 fully connected layers, for network intrusion detection with the UNSW-NB15 dataset~\cite{7348942}. 
	\end{itemize}
	
	%%% How Sections are organized
	This paper is organized as follows: Section~\ref{sec:hls_rtl} highlights the differences between HLS and RTL and the implied design and workflow trade-offs before Section~\ref{sec:rel_work} provides a literature review of related work. Section~\ref{sec:finn} briefly describes the FINN platform for generating HLS architectures for QNN inference. The architecture of the matrix vector compute unit and its RTL refinement are detailed in Section~\ref{sec:mvcu_arch}. The results of our HLS vs. RTL comparison are finally presented in Section~\ref{sec:res} before Section~\ref{sec:conc} concludes the paper.
	
	\section{High-Level Synthesis (HLS) vs. Register-Transfer Level (RTL) Design}
	\label{sec:hls_rtl}
	
	The design of digital microelectronic systems has evolved tremendously over the years. Starting with the manual placement of transistors, it eventually embraced hardware description languages (HDLs) as a major milestone boosting reuse and productivity. HDLs can describe a digital design at various abstraction levels, commonly identified as the behavioral, register-transfer, gate and switch levels. Typically, a combination of behavioral and register-transfer level is used for design entry with few instances of gate level design where detailed control is necessary. Conventionally, the term \emph{RTL design} is used as an umbrella subsuming the adjacent abstraction levels as used for typical design entry.
	
	RTL design entry was an important but by far not conclusive step towards a functional, software-like specification of hardware systems. Behavioral control structures, like \lstinline{if}- and \lstinline{case}-statements, and high-level logic and arithmetic operators raise the abstraction level significantly. However, designers must still schedule their computation into clock cycles explicitly and deal with low-level control details, such as the synchronicity of clock domains and the polarity of reset signals. Listing~\ref{lst:sv_dff} gives a small impression of this style of design entry modelling a D-Flip Flop with a synchronous active-low reset and an active-high enable.
	
	\begin{lstlisting}[float=t,xleftmargin=.3\textwidth,xrightmargin=.3\textwidth,frame=tb,caption=Simple D-Flip Flop in SystemVerilog, label={lst:sv_dff},language=Verilog,numbers=left]
		always_ff @(posedge clk) begin
		if(!rst_n)   Q <= 0;
		else if(en)  Q <= d;
		end
	\end{lstlisting}
	
	Three HDLs are in widespread use today: VHDL, Verilog, and SystemVerilog. SystemVerilog is the youngest among them and is closely related to Verilog. All of them require special skills in RTL design and are exclusively used by hardware engineers to implement designs on hardware platforms like application-specific integrated circuits (ASICs) and field-programmable gate arrays (FPGAs) and to build testbenches for these designs.
	
	In order to enable software engineers, research scientists and others to deploy and accelerate their applications on FPGAs without the need for writing RTL, high-level synthesis (HLS) tools were envisioned and created \cite{Nabi2017J}. These efforts yielded, for example, Maxeler's OpenSPL \cite{Pell_2012J},
	%\url{https://www.maxeler.com/openspl-announced/},
	Intel's OpenCL \cite{Czajkowski2012C} and Xilinx' Vitis unified software platform \cite{Vitis_HLS}. While Intel's HLS \cite{Intel_HLS} uses OpenCL, the Xilinx Vitis platform uses C++ as its design entry language. Pragma annotations, comparable to those used by OpenMP, guide the RTL synthesis by the HLS compiler. These pragma recipes and language restrictions, particularly the lack of dynamically allocated memory and late virtual function dispatch, still pose a learning and implementation challenge. This challenge can, however, be approached within familiar C++ terrain.
	
	HLS technology has, indeed, demonstrated its ability to make FPGA acceleration accessible to a wider audience. It has triggered a number of contributions using HLS in diverse fields, e.g., scientific high-performance computing (HPC) \cite{Nabi2019J,Muslim2017J} as well as machine learning applications like routing congestion prediction \cite{Zhao_2019C} and deep neural networks \cite{Noronha_2018C}. FINN \cite{Umuroglu_2017C,Blott2018J} is also an example for realizing accelerators for neural networks on FPGAs using C++ HLS. 
	
	The design entry in HLS comes with significant benefits. The time and effort to get an initial design up and running is significantly lower. Its functional validation can be conducted efficiently in software. The higher-level description is much more flexible and, hence, more amenable for a thorough design space exploration. Re-arranging loop nests and tuning the degree of parallelism are small changes that can be evaluated quickly. Finally, the quality of the hardware designs generated from HLS has been improving significantly and constantly over the recent years which led to meanwhile widespread adoption of this new form in descign entry.
	
	Designing on a higher functional abstraction level does, however, also incur costs. First of all, HLS produces some wrapping overhead for standardizing all control and communication interfaces of a design. Internally, HLS tools can brilliantly apply formalized knowledge about transforming and optimizing standard code structures. However, they have no intuition about exploitable application constraints and the consequences of custom control structures. They do not automatically push into optimizing congested parts of the design and regularly fail in scaling a design up or in meeting the expected timing. When a critical generated kernel violates physical design constraints, the available means and levers are much less clear. Cutting one or two LUT levels from a critical path that did not meet timing typically turns into a massive investigative endeavor, which cannot even be expected to be approached by the high-level audience of HLS users. Last but not least, current HLS compilers perform an additional compilation step with synthesis times that clearly grow superlinearly. The compilation of designs that somewhat fill modern high-end devices may take days. This may squander the gains of faster turnaround times in earlier design phases. In the end, the system implementation cost of an HLS application may suddenly surge when operating the tools close to the limits of the target platform. Particularly for such challenging and for settled use cases, RTL still poses as a worthy option.
	
	\section{Related Work}
	\label{sec:rel_work}
	As HLS design entry has become main stream, a number of research contributions for its analysis and evaluation have been made. Nane et al. \cite{Nane_2016J} present a survey and evaluation of FPGA HLS tools. In addition to evaluating various tools, they also describe different optimizations and problems being addressed by researchers. They identified benchmark-specific optimization and constraints that can enable HLS tools to significantly improve performance while also highlighting the differences between optimizations needed for hardware designs as compared to software designs. Furthermore, they also present an analysis of academic HLS tools against commercial ones.
	
	An overview of HLS techniques and tools has also been compiled by Coussy et al. \cite{Coussy_2009J} who outline the steps involved in an HLS design flow and compare it against an RTL design flow. A similar overview of a much larger set of tools was contributed by Meeus et al. \cite{Meeus_2012J}. They outline a number of challenges faced by HLS, such as application-specific tool flows, more optimization options for improving HLS designs, and design entry standardization asking for additional training for designers to adapt their applications for hardware platforms. Martin et al. \cite{Martin_2009J} divide HLS tools into  generations highlighting both their key features along with their shortcomings. According to the authors, the HLS of the early to late 2000's establishes the $3^{rd}$ generation comprising tools like Xilinx's AccelDSP, Synopsys Synplicity Synplify DSP among others. Although not documented, surely we are seeing a new generation of tools like Xilinx's VivadoHLS and VitisHLS and Intel's OpenCL-based HLS tool.
	
	In addition to these contributions towards the analysis of HLS, specific algorithms have been realized in HLS and compared against RTL counterparts. Homsirikamil et al. \cite{Homsi_2017C} present such an analysis for various cryptography algorithms. They use Xilinx's Vivado\,HLS and benchmark the algorithms on throughput and throughput/area metrics. Another relevant work is presented by Winterstein et al. \cite{Winterstein_2013C}. They analyze HLS for two types of implementations of the K-means algorithm, a dataflow and a recursive tree implementation. They demonstrate that dataflow architectures described using HLS achieve near RTL performance but recursive architectures do not.
	
	It is to be noted that FINN~\cite{Umuroglu_2017C,Blott2018J}, the base for the HLS vs. RTL comparison in this paper, in fact generates HLS dataflow architectures. This work performs an analysis on a large scale sweeping through the design space of a critical component of FINN networks. We implement the main compute unit of a neural network layer with varying design parameters, such as the input feature map size, kernel dimension, output feature map size, input and kernel word length, both from FINN's native HLS output and from a drop-in RTL description. We determine the relationship between the chosen design parameters and achieved quality metrics including throughput, resource utilization, end-to-end delay, and tool execution time.
	
	\section{FINN}
	\label{sec:finn}

	Xilinx provides an open-source framework called FINN~\cite{Umuroglu_2017C,Blott2018J} to generate highly specialized accelerators on FPGAs leveraging \emph{reduced-precision datatypes} and \emph{streaming dataflow} (DF) architectures. FINN customizes the hardware architectures to the specifics of a DNN topology and the exact datatypes used.
	Each layer is instantiated with its designated compute units in hardware. On-chip data streams interconnect the compute units to form the desired network topology. 
	The small and compact size of reduced-precision DNNs allows to store all parameters on the chip - off-chip memory access and with that potential memory bottlenecks are avoided.
	FINN produces synthesizable C++-based HLS code for describing the generated QNN accelerators. HLS was initially chosen for the hardware description of these highly customized architectures, as C++ templates can be used to parameterize canned designs for different datatypes, different layer types and different degrees of parallelism. Furthermore, it enables rapid implementation in comparison with handwritten RTL.
	In the following, we discuss the hardware architecture and the tool flow of FINN in separate sections.
	
	%Furthermore, FINN uses 1-bit values for all inputs and output where $0$ represents $-1$ and $1$ represents $+1$. This transforms a neural network into an XOR network as the basic multiplication between input and weight kernel is reduced to a XOR operation. 
	
	%In addition to this, FINN quantifies the peak performance of a BNN on an FPGA using a roofline model and optimizes the mapping of such BNNs onto these FPGAs by proposing an architecture and accelerator construction tool and presents a number of different prototypes to highlight the potential of BNNs on off-the-shelf FPGA platforms. 

	\subsection{Hardware Architecture}
	
	As mentioned above, the overall architecture realized by the FINN framework consists of multiple layers connected using a data-flow model.
	Popular layers such as convolutions are lowered to a matrix-matrix multiplication between the filter kernel and an activation matrix resulting from the expansion of the input feature map by the \texttt{im2col} operation. FINN performs this expansion on the fly using a sliding window moving across the input.
	As a result, each layer is a dedicated composition of sliding window unit (SWU) and matrix vector threshold units (MVU), with the MVU being the central compute block for convolutional layers, and similarly also for fully connected layers.
	This compute block, which represents the main compute engine in FINN designs, is the focus of this work. 
	It is parameterized in terms of number of input and output feature map channels, kernel dimensions, input feature map dimensions and input and kernel weight precision. 
	Furthermore, the degree of parallelism of a MVU can be specified through 2 parameters, namely the number of processing elements (PEs) with the width of singe-instruction multiple-data (SIMD) lanes which will be discussed in more detail in the next subsection on MVU architecture.

	\subsubsection{Matrix Vector Compute Unit Architecture}
	\label{sec:mvcu_arch}
	
	The matrix vector compute unit (MVU) is the main computation unit of each neural network layer's implementation in the FINN framework. While it also subsumes the output thresholding in this framework, this part of its functionality is not considered in this work as it only requires a few look-up tables (LUTs) for its realization \cite{Blott2018J}.
	
	As previously mentioned, in FINN, convolutions are lowered to matrix-matrix multiplications where the weights defining the convolution filters are packed into a matrix and the input image is also converted to a matrix by moving a sliding window across it \cite{Blott2018J}. This is known as general matrix multiply (GEMM) and quantized versions of it have been used to realize neural networks \cite{Umuroglu_2017Cc, gemmlowp}. 
	
\begin{figure}
		\centering
		\includegraphics[scale=0.5]{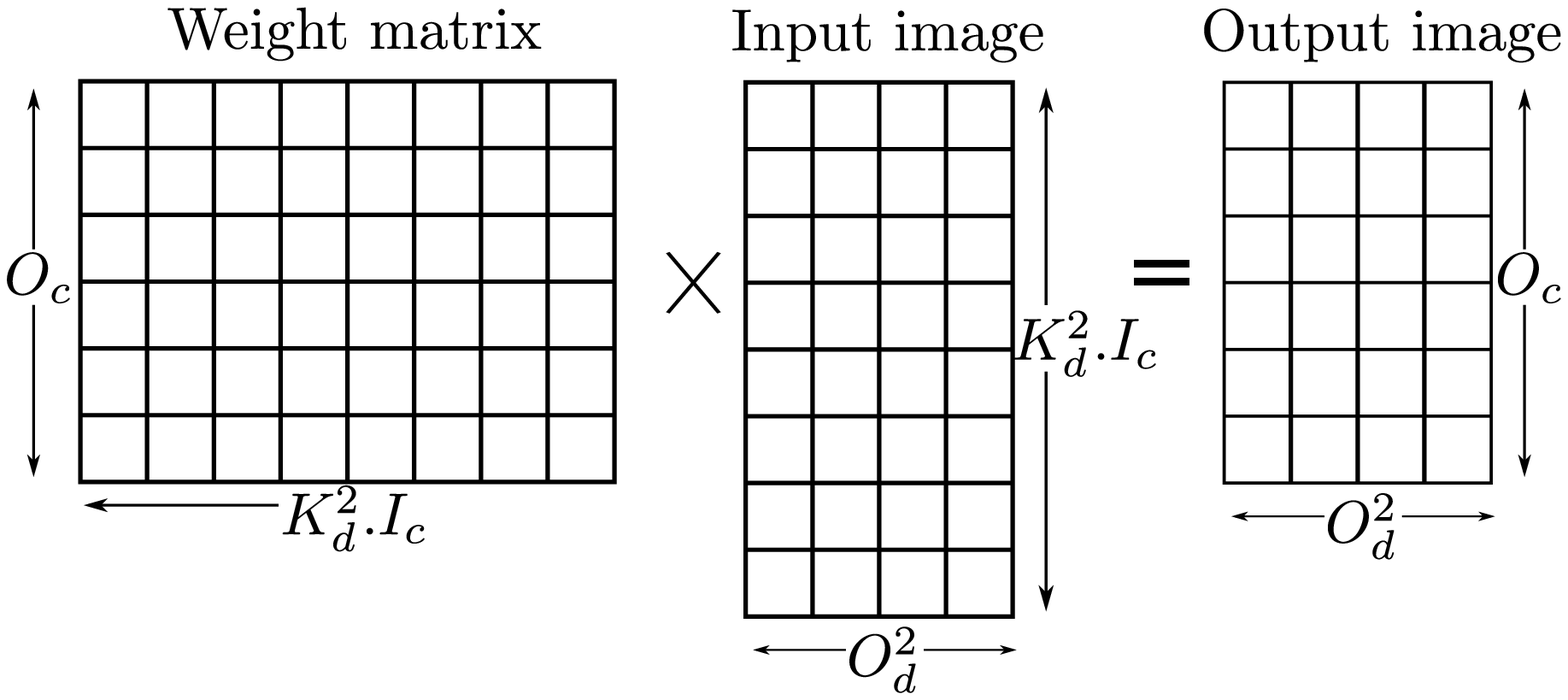}
		\caption{Input, weight and output matrix of convolution implemented as GEMM.}
		\label{fig:conv_matrix}
	\end{figure}

	The image matrix is generated by the sliding vector unit and then processed by the MVU along with the weight filter matrix. The image matrix has dimensions of $K_d^2.I_c \times O_d^2$, where $K_d, I_c$ and $O_d$ are the weight filter kernel dimensions, number of input feature map channels and dimensions of the output feature map, respectively. Similarly, the weight filter matrix has dimensions of $O_c \times K_d^2.I_c$, where $O_c$ is the number of channels in the output feature map. The output image matrix obtained after multiplying these two matrices have dimensions of $O_d^2 \times O_c$. This is also shown graphically in 
	Fig.~\ref{fig:conv_matrix}.
	
	Each vector of the input image matrix is streamed into the MVU along with the weight matrix to produce one output vector which is streamed to the next layer in the network. The AXI-Stream protocol is used for the communication between layers.
	
	\paragraph{MVU: Achieving Parallelism}
	\label{subsec:mvu_parallel}
	
\begin{figure}
		\centering
		\includegraphics[scale=0.3]{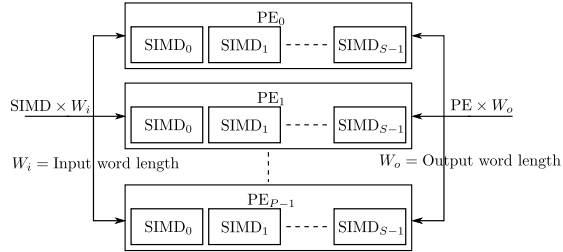}
		\caption{Processing elements (PEs) and SIMDs arrangement.}
		\label{fig:pe_simd}
	\end{figure}

	FPGAs allow for massive data parallelism and also the MVU can be parallelized in various dimensions with the bounds of the resources available in the target device. To achieve this, the MVU is divided into $P$ parallel processing elements (PEs) and $S$ single-instruction, multiple-data (SIMD) input lanes for each PE, as shown in Fig.~\ref{fig:pe_simd}. Here, each PE correspond to a hardware neuron and each SIMD to a hardware synapse \cite{Umuroglu_2017C}.
	
	For a fully parallel architecture, the number of PEs should equal the number of rows of the weight matrix. Thus, it means that each PE reads in a row of the weight matrix and multiplies it with the input image vector to produce an output vector. Within each PE, number of SIMD lanes will be equal to the number of columns in the weight matrix. This arrangement is graphically shown in Fig.~\ref{fig:pe_simd}. 
	
	However, resources in an FPGA are limited and a fully parallel implementation of the MVU may not be possible. Or conversely, throughput requirements are not high. Then, it is important to time-multiplex or fold the QNN onto fewer hardware resources. Time multiplexing is a general technique where only parts of the input are used in a given time instant for computation on a given hardware resource and the hardware resource is re-used for other time instances \cite{asad_frm2016J}.
	
	Consider an example where the weight matrix has a $4\times4$ dimension while the input image is a $4\times1$ vector. Let the number of PEs equal two with two SIMD lanes in each PE. Assume that the four elements of the input vector are $[x_0, x_1, x_2, x_3]$ and those of the weight matrix are:
	
	\begin{equation}
		Y = \begin{bmatrix}
			y_{00} & y_{01} & y_{02} & y_{03} \\
			y_{10} & y_{11} & y_{12} & y_{13} \\
			y_{20} & y_{21} & y_{22} & y_{23} \\
			y_{30} & y_{31} & y_{32} & y_{33} \\
		\end{bmatrix}
		\label{eq:wgt_mat_ex}
	\end{equation}
	
\begin{figure}
		\centering
		\includegraphics[scale=0.75]{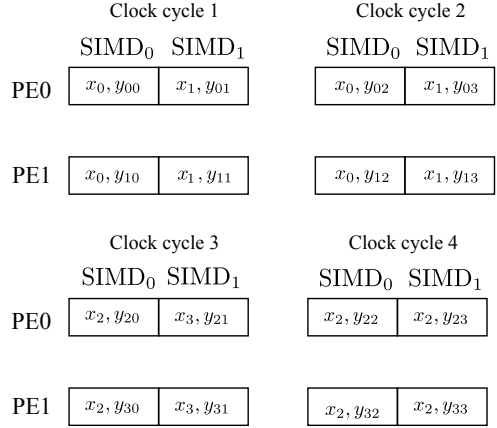}
		\caption{Mapping of input vector/weight matrix to each PE/SIMD.}
		\label{fig:map_fold}
	\end{figure}

	The mapping between the input vector/weight matrix to each PE/SIMD, on each clock cycle, is shown in Fig.~\ref{fig:map_fold}. It can be seen that elements of the input vector are re-used, thus necessitating the use of an input buffer. This results in a folded or time-multiplexed architecture whereas setting PE and SIMD to be one will result in a fully unfolded, serial architecture.
	
	\paragraph{Processing Element and SIMD Lanes}
	\label{subsec:mvu_pe}
	
	Each processing element consists of several SIMD lanes. This arrangement is shown in Fig.~\ref{fig:pe_simd}. The accumulator is only needed in case of folded architectures to accumulate the outputs of each clock cycle until the final output is computed.
	
\begin{figure}
		\centering
		\includegraphics[scale=0.75]{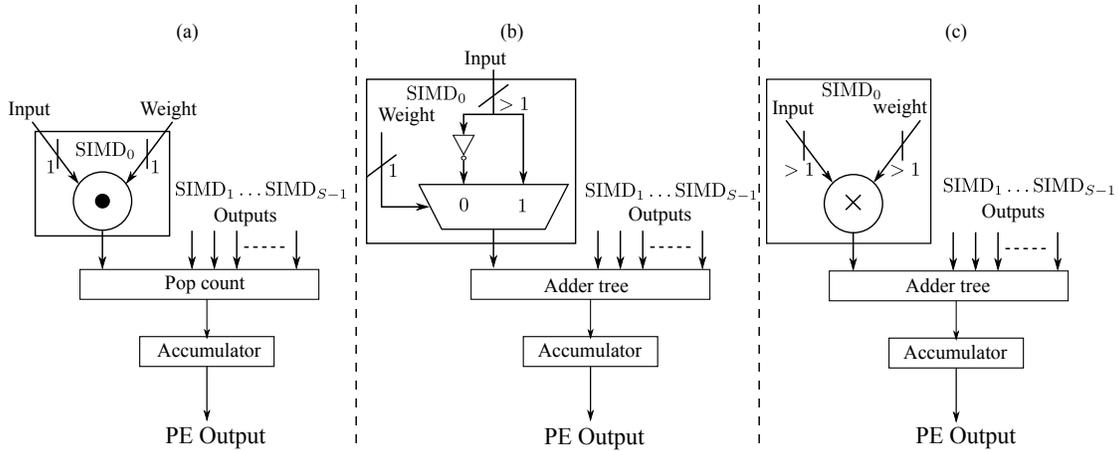}
		\caption{Types of SIMD components used. (a) XNOR, (b) Multiplexer based for binary inputs, where $0$ correspond to $-1$ and $1$ to $+1$ and (c) Standard multiplier for arbitrary precision inputs.}
		\label{fig:simd}
	\end{figure}

	Each SIMD lane essentially computes the product of the input vector element and weight matrix element, as also shown in Fig.~\ref{fig:map_fold}. FINN only used BNNs \cite{Umuroglu_2017C} which was then extended to binary weights and arbitrary precision input vectors in FINN-R \cite{Blott2018J}. Thus, the architecture of MVU supports three different types of SIMD lanes/elements for the different datatypes. The implementations are listed here and shown in Fig.~\ref{fig:simd} along with associated logic for adding the output of each SIMD and accumulator.
	
	\begin{itemize}
		\item XNOR followed by popcount
		\item Binary weights, interpreted as $\{\pm1\}$, followed by an adder tree
		\item Arbitrary precision weights and inputs, followed by an adder tree
	\end{itemize}
	
	The output of each SIMD lane in a PE can be added using a pop count, in case of binary inputs or a simple adder tree, as is used in this work, or even advanced compressor trees as proposed by Preußer \cite{preusser:2017} or Kumm and Kappauf \cite{kumm:2018}.

	\subsection{Tool flow}
	
	\begin{figure}
		\centering
		\includegraphics[scale=0.5]{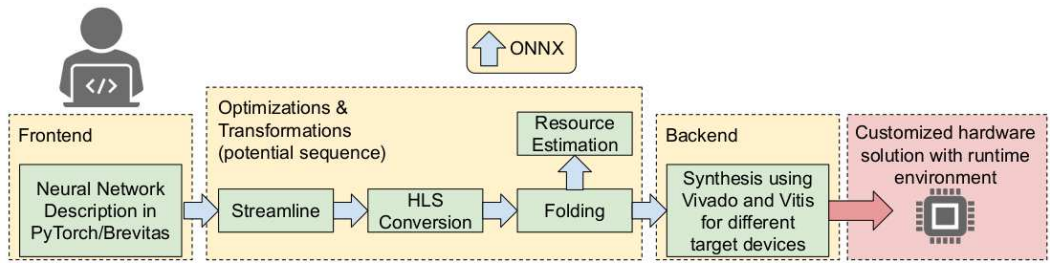}  
		\caption{FINN compiler flow.}
		\label{fig:finnflow}
	\end{figure}
	
	The FINN tool flow has a highly modular structure as shown in Figure~\ref{fig:finnflow}, which allows the user to interactively generate a specialized architecture for a specific DNN. 
	The framework provides a frontend, transformation and analysis passes and backends to explore the design space in terms of resource and throughput constraints.
	
	\paragraph{Frontend and Intermediate Representation} 
	The used training frontend is called Brevitas~\cite{brevitas2}.
	It is PyTorch library for quantization-aware training and enables training DNNs with weights and activations quantized down to a few bits, then exports the trained network into the intermediate representation (IR) used by the FINN compiler. 
	After training, the DNN model must be first converted into the IR of the FINN compiler. The frontend stage takes care of this by converting the PyTorch description into the IR, called FINN-ONNX. This IR is based on ONNX~\url{https://onnx.ai}, an open-source interchange format which uses a protobuf description to represent DNNs. The IR forms the input to the transformation and analysis passes.
	%It comes with several standard operators and allows the user to easily create their own operators to customize the model. The nodes represent layers and edges carry outputs from one layer to become inputs to another. The feature to customize the ONNX representation is used in the framework to add application-specific nodes and attributes. Each node is tagged with the quantization of its inputs, parameters and outputs to facilitate quantization-aware optimizations and the mapping to backend primitives optimized for quantized computation.  
	
	\paragraph{Transformation and Analysis Passes} 
	The \emph{transformation and analysis passes} help to generate an efficient representation of the DNN. 
	For this, the \emph{FINN compiler} performs graph transformation and analysis passes, which analyze and change the IR of the model. 
	%A pass is a function that takes the IR graph as its input and either (a) \emph{transforms} the DNN by looking for a certain pattern, modifying the graph in a specific manner before yielding it as its outputs, or (b) \emph{analyzes} the DNN to produce metadata about its properties. To bring the model into a representation, from which code can be produced, various transformations must be applied. 
	In this part of the compiler flow, synthesizable HLS descriptions are generated for the various layers (see \emph{Lowering and Conversion to HLS Layers}) and the degree of parallelization is determined, which is essential to meet specific resource or throughput constraints (see \emph{Folding and Resource Estimation} below).
	
	%The main transformations involved are summarized below.
	%\paragraph{Streamlining} Although the PyTorch description of the network is mostly quantized, it may still contain some floating-point operations from e.g. preprocessing, channelwise scaling or batchnorm layers. In order to generate a hardware accelerator from the model, these floating-point operations are absorbed into multi-level thresholds creating a functionally identical network of integer operations. The corresponding transformation is called \emph{streamlining} and was described by Umuroglu and Jahre~\cite{DBLP:journals/corr/abs-1709-04060}. During streamlining, floating-point operations are moved next to each other, collapsed into a single operation and absorbed into subsequent multi-thresholding nodes.
	
	\paragraph{Lowering and Conversion to HLS Layers} High-level operations in the graph are \emph{lowered} to simpler operators implemented by the FINN hardware library. As mentioned before, convolutions are lowered to a sliding window node followed by a MVU node. In the resulting graph, each node corresponds to a Vivado HLS C++ function call, for which an IP block can be generated using Vivado. The resources utilized by each hardware building block can be controlled through specific attributes passed from FINN to Vivado. For example, multiplications can be performed using LUTs or DSP blocks, and parameters can be stored in distributed, Block, or Ultra RAM. The main resource adjustment happens in the \emph{Folding and Resource Estimation} pass.
	
	\paragraph{Folding and Resource Estimation} The folding process assigns compute resources to each layer to obtain the desired throughput within a balanced pipeline. This process in essence determines values for PE and SIMD. 
	%\CHANGE{The underlying Vivado HLS library that provides the hardware building blocks for FINN supports controlling the degree of parallelism along the P and S dimensions from Figure \ref{fig:parallelism_sw_view}, called PE and SIMD respectively in FINN. To enable per-layer specialization without reconfiguration and minimize latency, FINN creates dedicated per-layer hardware interconnected with FIFO channels, thus the outermost loop across $L$ layers is always fully pipelined.}
	Once the folding is specified, resource estimates can be produced for each node. 
	
	There are several ways to estimate the resources. Even before IP blocks are generated from the HLS layers, an estimate of the resources per layer can be made by using analytical models based on the concepts from the FINN-R paper~\cite{Blott2018J}. 
	Estimations can also be extracted from Vivado HLS after IP generation, though these results are still estimations that may differ from the resource usage of the final implementation due to synthesis optimizations.
	%\CHANGE{Finally, FINN allows high-quality resource estimates to be obtained through out-of-context synthesis of nodes, at the expense of runtime. The most suitable estimation method depends on the context (e.g. prototype vs production phases).}
	
	\paragraph{Backends} Finally, backends are responsible for consuming the IR graph and translating these into RTL descriptions and backend-specific information to create a deployment package.
	FINN support implementation as a standalone Vivado IP core or integrated into various shells, such as the ones available for Xilinx Alveo boards and PYNQ embedded platforms.

	\section{The MVU RTL Implementation}
	\label{subsec:mvu_rtl}
	
	\subsection{Overall Architecture}
	\label{subsec:mvu_overall}
	
	Realizing the MVU in HLS only requires defining the matrix vector multiplication, partitioning the matrices and internal arrays, and defining the pipelining. The control logic required to synchronize all these operations, including the AXI stream I/O protocol, is handled by the HLS framework. For RTL, however, one needs to define the complete control logic manually.
	
	\begin{figure}
		\centering
		\includegraphics[scale=0.4]{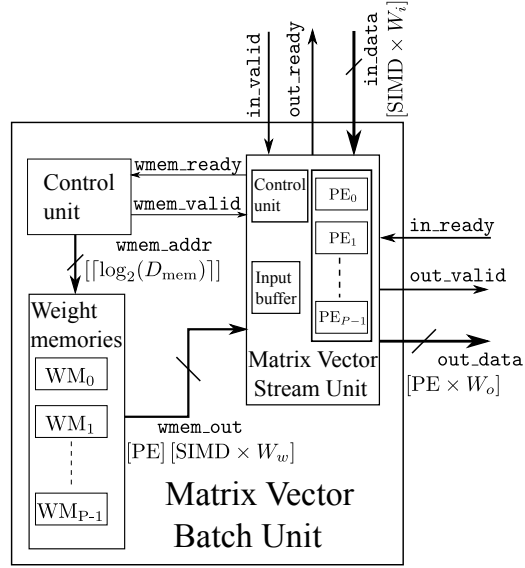}
		\caption{Overall architecture of the MVU batch and stream units.}
		\label{fig:top_module}
	\end{figure}
	
	For the RTL implementation, the overall architecture of the MVU, as shown in  Fig.~\ref{fig:top_module}, was divided into two modules, one containing the other. The top-level module is referred to as \emph{MVU batch} and incorporates burned-in weight memory, a control unit, which is responsible for sequencing the stream of weights to the compute in the stream unit, as well as the stream unit \emph{MVU stream} itself, which is the second module. The latter implements the main computation partitioned along the two dimensions of PEs and SIMD accumulations.
	Both slices of weights and input data are streamed in parallel into the unit. This partition is inherited from FINN~\cite{finn}.

	In order to understand the sequencing and dimensions to parallelize, it is important to consider the layout of the weight memory. The depth of each weight memory is given by:
	\begin{equation}
		D_{\text{mem}} = \frac{K_d^2 \times I_c \times O_c}{\text{SIMD} \times \text{PE}}
		\label{eq:wmem}
	\end{equation}
	where $K_d, I_c$ and $O_c$ have their usual meanings as defined earlier. The word size of the data stored in these memories is $SIMD \times B_w$, where $B_w$ is the weight precision. Since each PE is responsible for one or more rows of the weight matrix, it is served by a dedicated weight memory resulting in PE memory instances.

	\subsection{MVU Batch Unit}
	\label{subsec:mvu_batch}
	
	The MVU batch unit, as shown in Fig.~\ref{fig:top_module}, comprises a small \emph{control unit} for managing the reads from these weight memories, whose contents are initialized offline and are also part of the MVU batch unit. The MVU batch unit then contains the main compute unit shown as the matrix vector stream unit in Fig.~\ref{fig:top_module}. 
	
	\subsection{MVU Stream Unit}
	
	The stream unit performs the  main computations. It encapsulates the PEs and SIMDs of Fig.~\ref{fig:pe_simd}. The stream unit also has a separate control unit which deals with the AXI stream protocol and handles the reading and writing of the input buffer. The design is generic both in terms of PE and SIMD partitioning and in terms of data precision by leveraging appropriate SystemVerilog \lstinline|generate| loops and conditional instantiations to select from the implementation alternatives introduced by Fig.~\ref{fig:simd}.

	\subsubsection{AXI Stream Protocol} 
	\label{subsec:mvu_axi}
	
	The AXI stream protocol \cite{axis} defines a standard interface for exchanging data between connected components. It follows a master/slave design with the slave accepting data from the master. From all its interface signals, this work only uses the five tabulated and described in Tab.~\ref{tab:axi_signals}.
	
	\begin{table}
		\centering
		\caption{AXI interface signals used in this work.}
		\label{tab:axi_signals}
		\begin{tabular}{ll}\toprule
			\lstinline|ACLK|    &  The global clock signal\\
			\lstinline|ARESETn| & The global active low asynchronous reset signal\\
			\lstinline|TVALID|  & Signal to indicate master is driving a valid output\\
			\lstinline|TREADY|  & Signal to indicate the slave can accept a valid input\\
			\lstinline|TDATA|   & The main data signal being exchanged\\\bottomrule
		\end{tabular}
	\end{table}
	
	Data is only transmitted when both handshaking signals, i.e. \lstinline|TVALID| and \lstinline|TREADY|, are asserted. By deasserting \lstinline+TREADY+, the receiving slave is able to assert \emph{backpressure} as a means of flow control. In the case of FINN, this enables a slower layer that is not ready to accept new data to pace its upstream data source. Other flow scenarios include the intermittent availability of data from the preceding layer and the intermittent assertion of the ready signal by the succeeding layer. All this necessitated the use of a finite state machine to synchronize the data transfer to and from the stream unit. Since data coming from a preceding layer is written to the input buffer and backpressure requires either to stop writing to the input buffer or stop reading from it (in case of a folded architecture), access to the input buffer is also handled by the state machine.
	
	\subsubsection{Finite State Machine}
	\label{subsec:mvu_fsm}
	
	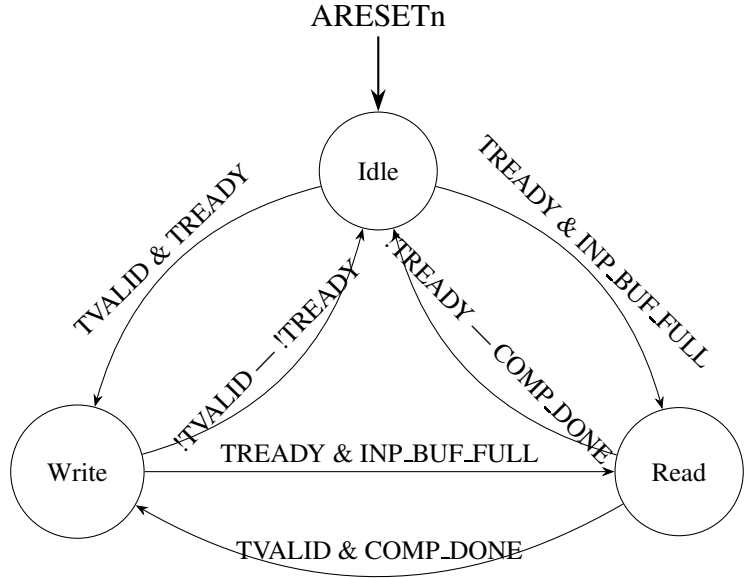
\begin{figure}
		\centering
		\tikzstyle{line} = [draw, -latex']
		\begin{tikzpicture}[node distance = 2.5cm,
			auto, initial text = {ARESETn},
			every initial by arrow/.style = {font = \large, thick,-{Stealth[length=3mm,width=2mm]}}
			]
			% \footnotesize
			%%% Placing the Nodes
			\node[state, inner sep=10pt, initial above, initial distance=1cm] (idle) {Idle};
			\node[state, inner sep=10pt] at (-4,-4) (write) {Write};
			\node[state, inner sep=10pt] at (4,-4) (read) {Read}; 
			
			% %%% Drawing Edges
			\draw (idle) edge[below, bend right, -Stealth] node[sloped, anchor=center, above] {TVALID \& TREADY} (write);
			\draw (write) edge[above, bend right, -Stealth] node[sloped, anchor=center, above=0.1cm] {!TVALID | !TREADY} (idle);
			\draw (idle) edge[below, bend left, -Stealth] node[sloped, anchor=center, above] {TREADY \& INP\_BUF\_FULL} (read);
			\draw (read) edge[above, bend left, -Stealth] node[sloped, anchor=center, above=0.25cm] {!TREADY | COMP\_DONE} (idle);
			\draw (write) edge[above, -Stealth] node[sloped, anchor=center, above] {TREADY \& INP\_BUF\_FULL} (read);
			\draw (read) edge[bend left, below, -Stealth] node[sloped, anchor=center, above=0.1cm] {TVALID \& COMP\_DONE} (write);
			
		\end{tikzpicture}
		\caption{State diagram of the finite state machine for controlling the MVU stream unit.}
		\label{fig:fsm}
	\end{figure}
	
	% \TODO{indicate where the FSM sits}
	The control that is part of the stream unit implements a three-state Mealy machine is implemented to synchronize the data movement between adjacent layers and the internal data processing. Its abstract representation is shown in Fig.~\ref{fig:fsm}. It identifies the states and their transitions. The idle state is the default state assumed at the start, during back-pressure and whenever no input is available from the preceding layer. The internal signals \lstinline|COMP_DONE| and \lstinline|INP_BUF_FULL| indicate the completion of the computation for one streamed input sequence and the fill up of the input buffer, respectively.
	As shown, the state machine transitions from the idle state to the write state when new input data is available. It remains there while data is available and the input buffer is not yet filled up. Already when filling the input buffer, the written input data is presented to the PEs so that they can start or proceed their operation. When the input buffer is filled, the state machine transitions to the read state where the buffered data is re-used for their processing with other rows of the weight matrix as illustrated by Fig.~\ref{fig:map_fold}. During both reading and writing, the computation can be stalled due to a number of reasons, which then cause the state machine to transition back to the idle state.
	
	Instead of halting the computation immediately upon back-pressure, the computation is allowed to proceed for a few cycles while a small temporary FIFO buffer captures the produced output. This decouples the production of outputs in bursts by the parallel PEs from the consumption of inputs in bursts by the parallel SIMD units of the next layer so that a smoother dataflow with more effective parallelism is achieved. In the case of unavailable data from the preceding layer, the computation must be halted.
	
	\section{Results}
	\label{sec:res}
	
	In this section, we evaluate the RTL and HLS implementation alternatives of the MVU with respect to different performance metrics. In Section~\ref{subsec:res_util}, we look at the resource utilization in terms of look-up table (LUT) and flip-flop (FF) counts and in terms of block RAM (BRAM) usage. We also report the total number of execution cycles, which are a measure of how long each implementation takes to process the same number of inputs.
	In Section~\ref{subsec:res_delay}, we analyze the critical path delay while in Section~\ref{subsec:syntimes}, we report the impact on synthesis time.
	Finally, in Section~\ref{subsec:res_app}, we look at a complete application of network intrusion detection. It uses a multi-layer perceptron (MLP) network composed of four fully connected layers \cite{Umuroglu_2020C}. We start with describing the experimental setup.
	
	\subsection{Experimental Setup}
	\label{subsec:experimental}

	% listed below: \TODO{Tom: I would prefer a table float. In fact, it seems that Tab.~\ref{tab:table_analysis} already does the job.}
	% \begin{itemize}
		%     \item Number of input feature map (IFM) channels
		%     \item Number of output feature map (OFM) channels
		%     \item IFM dimensions
		%     \item Kernel dimensions
		%     \item Number of processing elements (PEs)
		%     \item Number of SIMD elements per PE
		% \end{itemize}
	
	We analyse both RTL and HLS designs based on the network and design parameters. A list of these parameters are given in Table~\ref{tab:table_analysis}. In addition to these parameters, the MVU was synthesized with the three different types of SIMD elements shown in Fig.~\ref{fig:simd}, which will be referred to in this section as XNOR, binary weights and standard implementation. In the case of the standard implementation, we four as the precision for inputs and weights, which is in line with the word sizes typically used in the FINN framework \cite{Blott2018J}. 
	
	For analysis, we sweep through each of these parameters while keeping the others constant. Different configurations for the analysis are shown in Table~\ref{tab:table_analysis}. A star (``*'') in the place of a configuration parameter identifies the one being varied. For example, in configuration $1$, the number of IFM channels is modified while keeping all other parameters at the named constants.
	
	\begin{table}[!t]
		\centering
		\caption{Layer and implementation parameters for analysis.}
		\label{tab:table_analysis}
		\begin{tabular}{lccccccc}
			\toprule
			Configuration & 1 & 2 & 3 & 4 & 5 & 6 \\
			\midrule
			Num. of input feature map (IFM) channels & * & 64 & 64 & 64 & 64 & 64\\
			IFM dimensions & 32 & * & 32 & 32 & 8 & 8\\
			Num. of output feature map (OFM) channels & 64 & 64 & * & 64 & 64 & 64\\
			Kernel dimensions & 4 & 4 & 4 & * & 4 &4\\
			Num. of processing elements (PEs) & 2 & 32 & 2 & 32 & * & 64\\
			Num. of SIMD elements per PE (SIMDs) & 2 & 32 & 2 & 32 & 64 & *\\
			\bottomrule
		\end{tabular}
	\end{table}
	
	All the presented results were generated using the Xilinx Vivado and Vivado HLS tools for RTL and HLS, respectively. The targeted FPGA was a Xilinx's Zynq-7000 SoC, the XC7Z020-1CLG400C device of the Pynq-Z1 board. A number of designs, both HLS- and RTL-based, were tested on the board to allow the designs to go through the complete implementation and validation cycle. However, the reported results are the estimates obtained directly after the out-of-context (OOC) synthesis of the corresponding design since the MVU implementations are meant to be part of an enclosing top-level design. The OOC design flow allows units to be synthesized and analyzed independently of a concrete top-level design \cite{xil_ooc}. All inputs, outputs and clocks were properly constrained for the synthesis. The clock was typically constrained to a period of $5\,ns$, only to be increased to $10\,ns$ in case either HLS or RTL fail to meet the tighter target constraint. Finally, the reported synthesis times measure the complete processing of the design sources all the way to obtaining the synthesized netlist. In the case of HLS, this comprises both HLS and RTL synthesis. The number of PEs and SIMDs/PE are kept low when we modify the IFM and OFM channels so that we can realize circuits with with values ranging between $2$ to $64$ for these two parameters.
	
	\subsection{Resource Utilization}
	\label{subsec:res_util}
	
	\subsubsection{Look-up tables (LUTs) and Flip-flops (FFs)}
	\label{subsubsec:res_lutff}
	
	Look-up tables (LUTs) form the core of the general FPGA fabric used to implement digital logic. They can also serve as small distributed memories, which is very useful for implementing shallow and narrow memory efficiently. While synthesizing the MVU architecture, the choice whether to utilize block RAMs (BRAMs) or LUTs for memory was left to the synthesizer. The flip-flops are used as pipelining registers.
	
	% \TODO{this paragraph is a bit out of context here? perhaps better under experimental setup?}
	
	\begin{figure}[!t]
		\centering
		\includegraphics[scale=0.5]{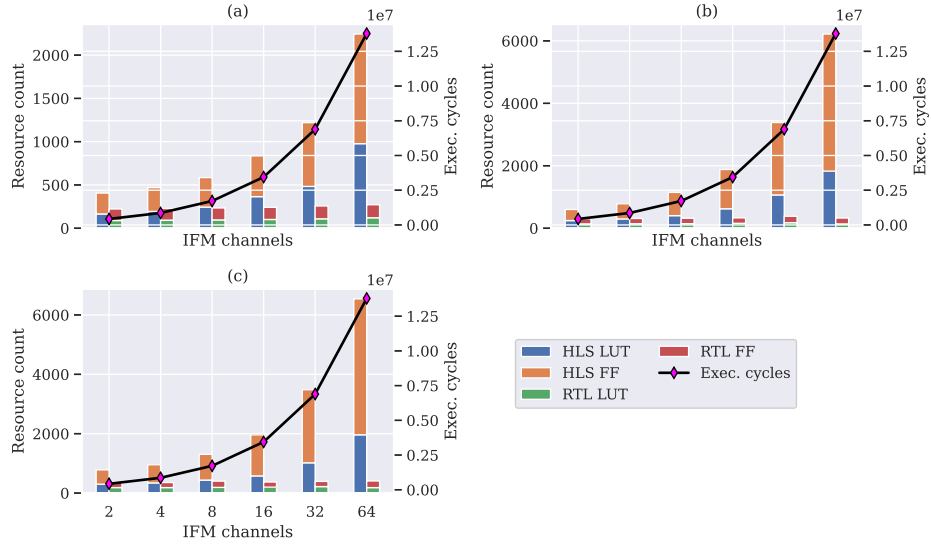}
		\caption{Resource utilization and latency, in terms of look-up tables (LUTs) and flip-flips (FFs), for HLS and RTL when varying number of input feature map channels with (a) 1-bit precision, (b) binary weights and (c) 4-bit precision.}
		\label{fig:res_util_ifmch_final}
	\end{figure}
	
	To analyse LUT and FF utilization, a number of parameters were modified as shown in Table~\ref{tab:table_analysis}, one at a time, while others were held constant when measuring their impact on resource utilization. 
	The number of input feature map channels (IFM channels) indicates the number of channels of the input image where each image has a certain dimension, in this case $32\times 32$. 
	The resource count in terms of LUTs and flip-flops is shown in Fig.~\ref{fig:res_util_ifmch_final} for the three types of SIMD elements shown earlier.
	
\begin{figure}[!p]
		\centering
		\includegraphics[scale=0.5]{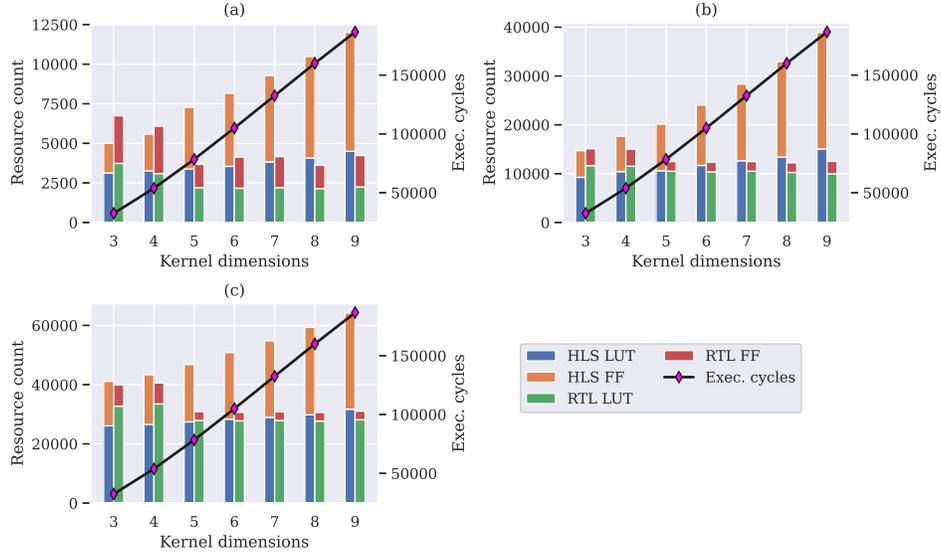}
		\caption{Resource utilization and latency, in terms of look-up tables (LUTs) and flip-flips (FFs), for HLS and RTL when varying kernel dimension with (a) 1-bit precision, (b) binary weights and (c) 4-bit precision.}
		\label{fig:res_util_kdim_final}
	\end{figure}

	As can be observed, the number of IFM channels does not impact the core architecture of the MVU as that is determined by the number of PEs and SIMDs. This is reflected in the total resource count usage by RTL remaining constant as we vary the values of IFM channels from $2$ -- $64$ and also in the increase in the total number of execution cycles indicating the same design is being re-used for more inputs. However, the depth of the input buffer, which is given by $\frac{K_d^2\times I_c}{\text{SIMD}}$, increases with a growing number of IFM channels. Input data to the PEs and SIMDs either comes directly from the input stream or is provided by the input buffer and with an increase in the depth of input buffer, HLS designs result in a more complex multiplexer architecture than RTL, thus affecting the total resource usage significantly. A similar effect is observed in Fig.~\ref{fig:res_util_kdim_final} when we explore different values of the kernel dimension, in the range $3$ -- $9$, which also affects the buffer length.
	
	\begin{figure}[!p]
		\centering
		\includegraphics[scale=0.5]{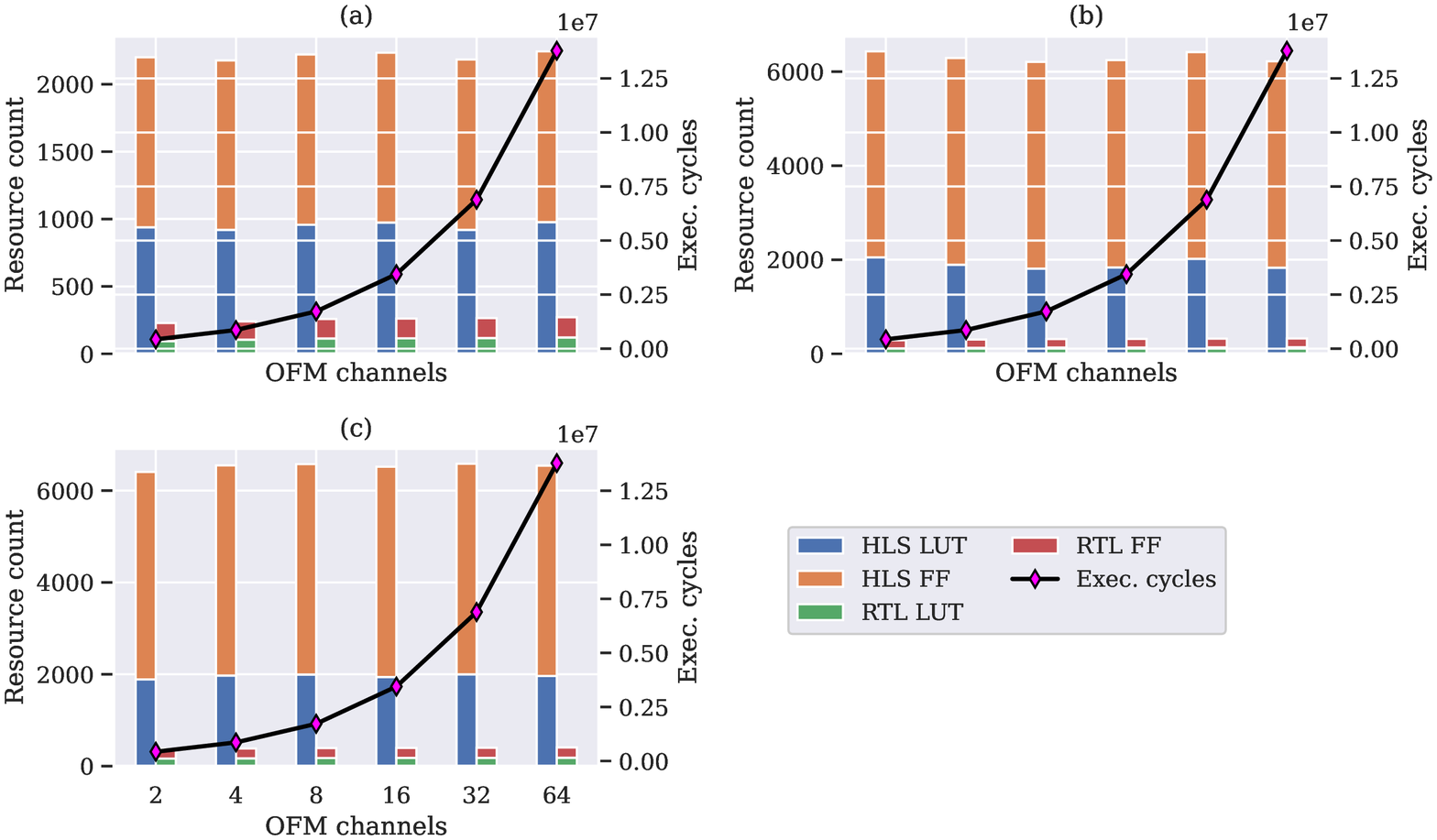}
		\caption{Resource utilization and latency, in terms of look-up tables (LUTs) and flip-flips (FFs), for HLS and RTL when varying number of output feature map channels with (a) 1-bit precision, (b) binary weights and (c) 4-bit precision.}
		\label{fig:res_util_ofmch_final}
	\end{figure}
	
	\begin{figure}[!p]
		\centering
		\includegraphics[scale=0.562]{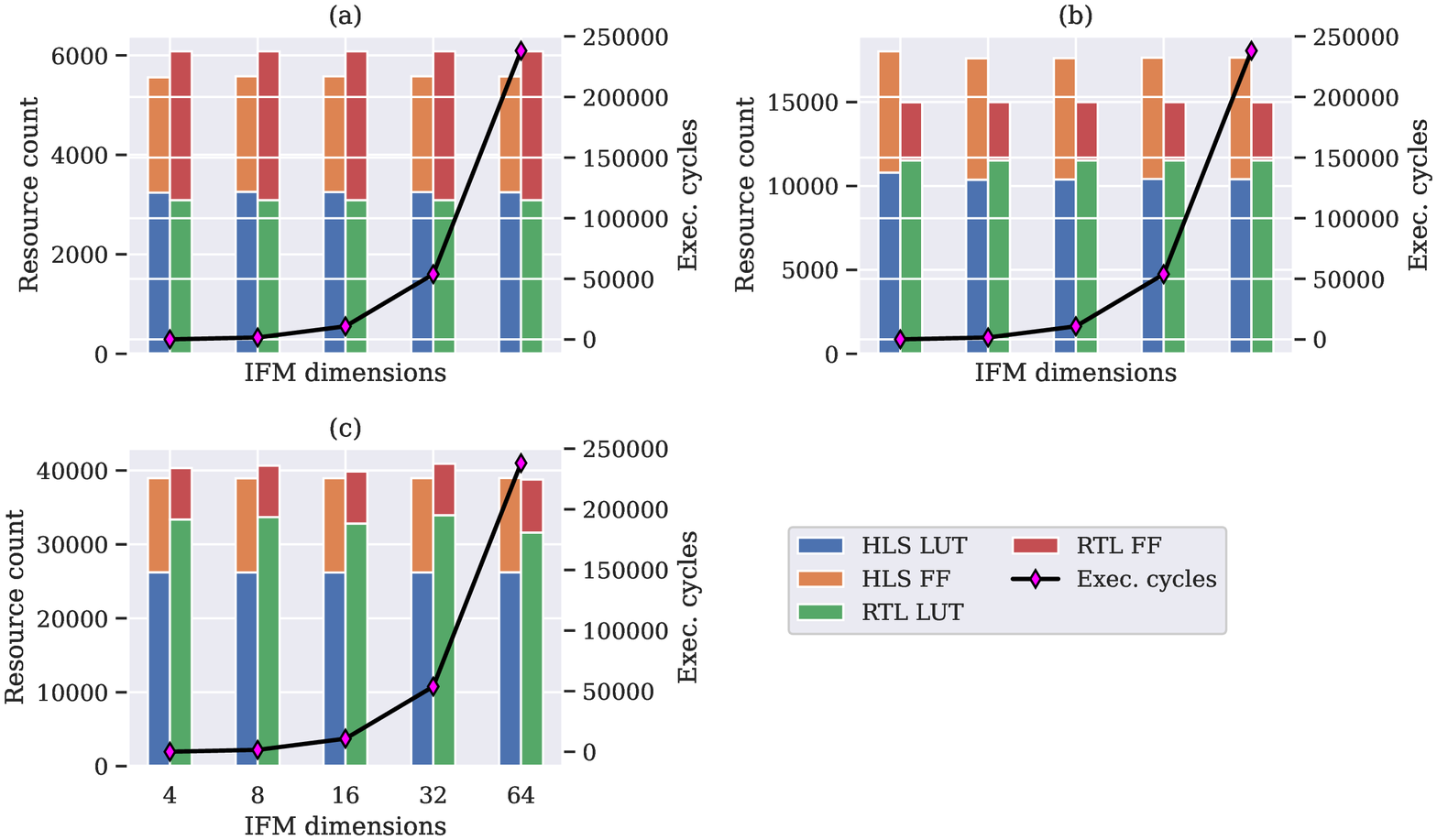}
		\caption{Resource utilization and latency, in terms of look-up tables (LUTs) and flip-flips (FFs),  for HLS and RTL when varying input feature map dimension with (a) 1-bit precision, (b) binary weights and (c) 4-bit precision.}
		\label{fig:res_util_ifmdim_final}
	\end{figure}
	
	\begin{figure}[!p]
		\centering
		\includegraphics[scale=0.562]{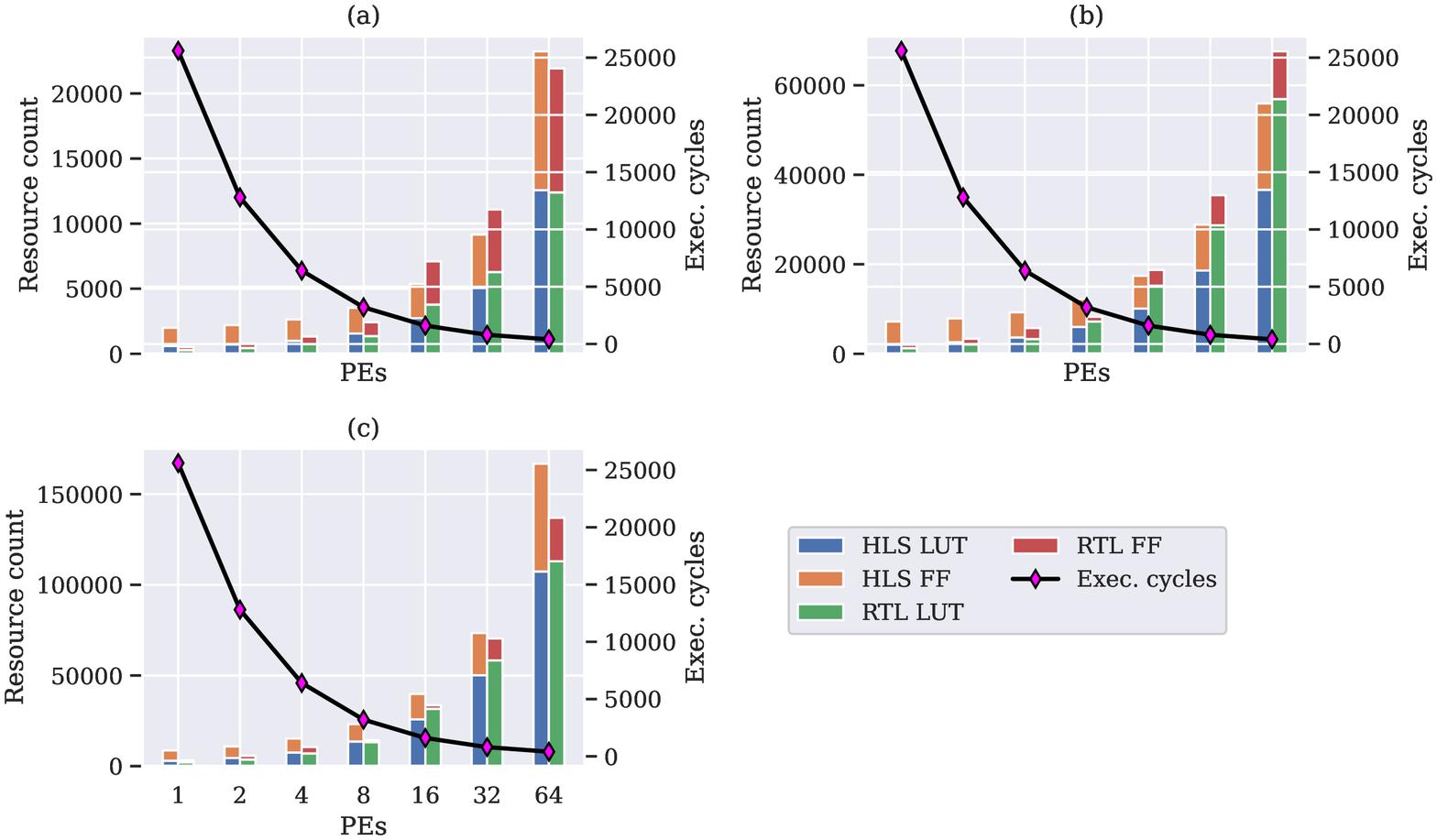}
		\caption{Resource utilization and latency, in terms of look-up tables (LUTs) and flip-flips (FFs),  for HLS and RTL when varying the number of processing elements (PEs) with (a) 1-bit precision, (b) binary weights and (c) 4-bit precision.}
		\label{fig:res_util_pe_final}
	\end{figure}
	
	\afterpage{\clearpage}
	
	However, when sweeping through the number of output channels (number of channels in the output activation), this increase is not seen since the buffer length and other associated logic remains the same. The only thing affected is the number of clock cycles needed to process all the inputs which is manifested in execution time. This is visualized in Fig.~\ref{fig:res_util_ofmch_final}.
	
	\begin{figure}[!t]
		\centering
		\includegraphics[scale=0.575]{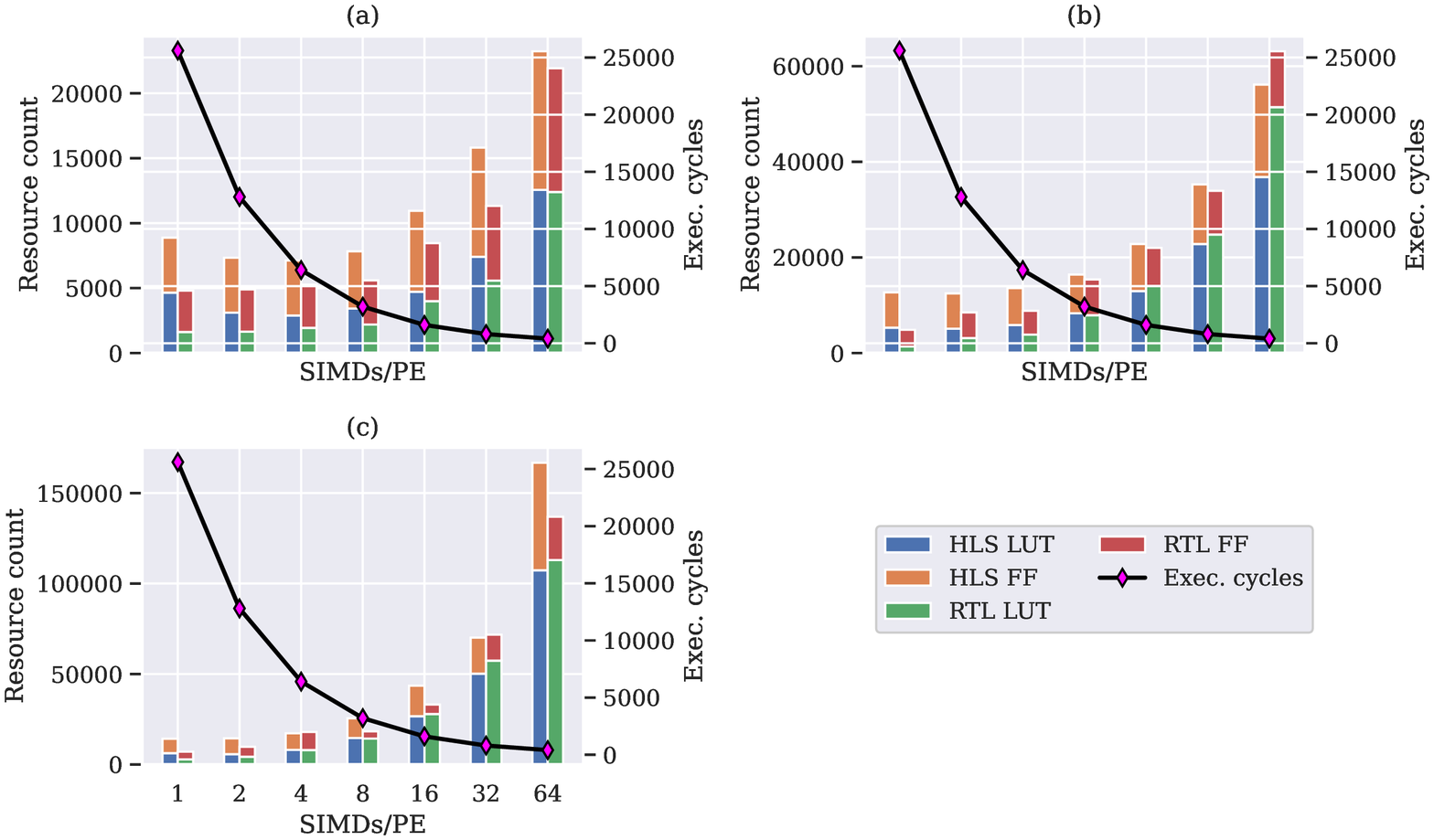}
		\caption{Resource utilization and latency, in terms of look-up tables (LUTs) and flip-flips (FFs), for HLS and RTL when varying the number of SIMD elements per PE with (a) 1-bit precision, (b) binary weights and (c) 4-bit precision.}
		\label{fig:res_util_simd_final}
	\end{figure}

	The core MVU used for Figs.~\ref{fig:res_util_ifmch_final}, \ref{fig:res_util_kdim_final} and \ref{fig:res_util_ofmch_final} is small, only consisting of two PEs and two SIMDs. For small designs, HLS tends to use more resources as compared to RTL, indicating an already large generated basic control logic in the implemented design. However, the amortization of this logic improves as the core MVU design is expanded by the increase in the number of PEs and SIMDs. This is highlighted in Fig.~\ref{fig:res_util_ifmdim_final} where the IFM dimensions are swept from $4$ to $16$ in steps of powers of $2$. Here, the number of PEs and SIMDs is $32$, indicating a large core MVU design. There is little difference in the resource usage by HLS and RTL and since the IFM dimensions do not impact the design complexity (only results in an increase in the execution cycles), the resource usage remains fairly consistent across the range of values.
	
	\begin{figure}
		\centering
		\includegraphics[scale=0.65]{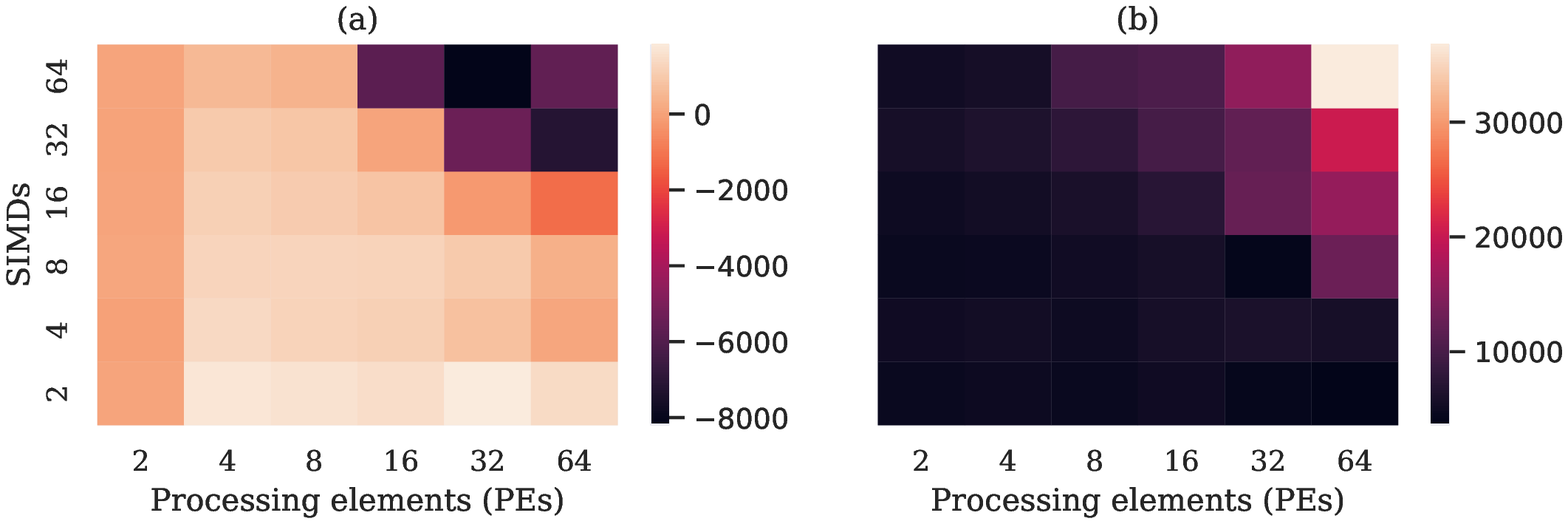}
		\caption{Heat map of difference in resource utilization between HLS and RTL in terms of (a) LUTs and (b) FFs when changing the number of processing elements (PEs) and SIMD lanes (SIMDs), for $4$-bit inputs.}
		\label{fig:pe_simd_heatmap}
	\end{figure}

	In order to see how HLS and RTL resource usage converge, we increase the number of PEs and SIMDs separately, again keeping the other dimensions constant.  Fig.~\ref{fig:res_util_pe_final} shows how the difference between HLS and RTL shrinks with increasing number of PEs. A similar effect is shown in Fig.~\ref{fig:res_util_simd_final} for increasing the number of SIMDs. In order to properly illustrate this relationship, Fig.~\ref{fig:pe_simd_heatmap} shows a heat map of the difference between resource utilization of HLS and RTL as PEs and SIMDs/PE are increased, with positive values showing RTL using fewer resources and negative ones otherwise.

	\begin{table}
		\centering
		\caption{Configuration parameters for larger designs with an increasing number of IFM channels.}
		\label{tab:table_large_design}
		\begin{tabular}{lccc}
			\toprule
			Configuration & 0 & 1 & 2 \\
			\midrule
			IFM channels & 16 & 32 & 64 \\
			IFM dimensions & 16 & 16 & 16 \\
			OFM channels & 16 & 16 & 16 \\
			Kernel dimensions & 4 & 4 & 4\\
			Weight precision & 4 & 4 & 4\\
			Input precision & 4 & 4 & 4 \\
			PE & 16 & 16 & 16\\
			SIMD & 16 & 16 & 16\\
			\bottomrule
		\end{tabular}
	\end{table}

	The heat map for the LUTs in Fig.~\ref{fig:pe_simd_heatmap}(a) shows that RTL uses less LUTs as compared to HLS for smaller designs and as designs are made larger by increasing the number of PEs and SIMDs, the LUT usage by HLS design is lower as compared to RTL design. However, the flip-flop usage by HLS is always more than RTL for all designs.
	
	\begin{table}
		\centering
		\caption{Resource utilization for configurations given in Table~\ref{tab:table_large_design}}
		\label{tab:res_large_design}
		\begin{tabular}{c|cc|cc}
			\toprule
			\multirow{2}{*}{Config.} & \multicolumn{2}{c|}{LUTs} & \multicolumn{2}{c}{FFs}\\
			& HLS & RTL & HLS & RTL \\
			\midrule
			Config. \#0 &  7528 &  7572 & 8400 & 5838\\
			Config. \#1 &  7354 & 7599 & 7560 & 5857\\
			Config. \#2 &  7919 & 8102 & 9634 & 5659\\
			\bottomrule
		\end{tabular}
	\end{table}
	
	\begin{figure}
		\centering
		\includegraphics[scale=0.6]{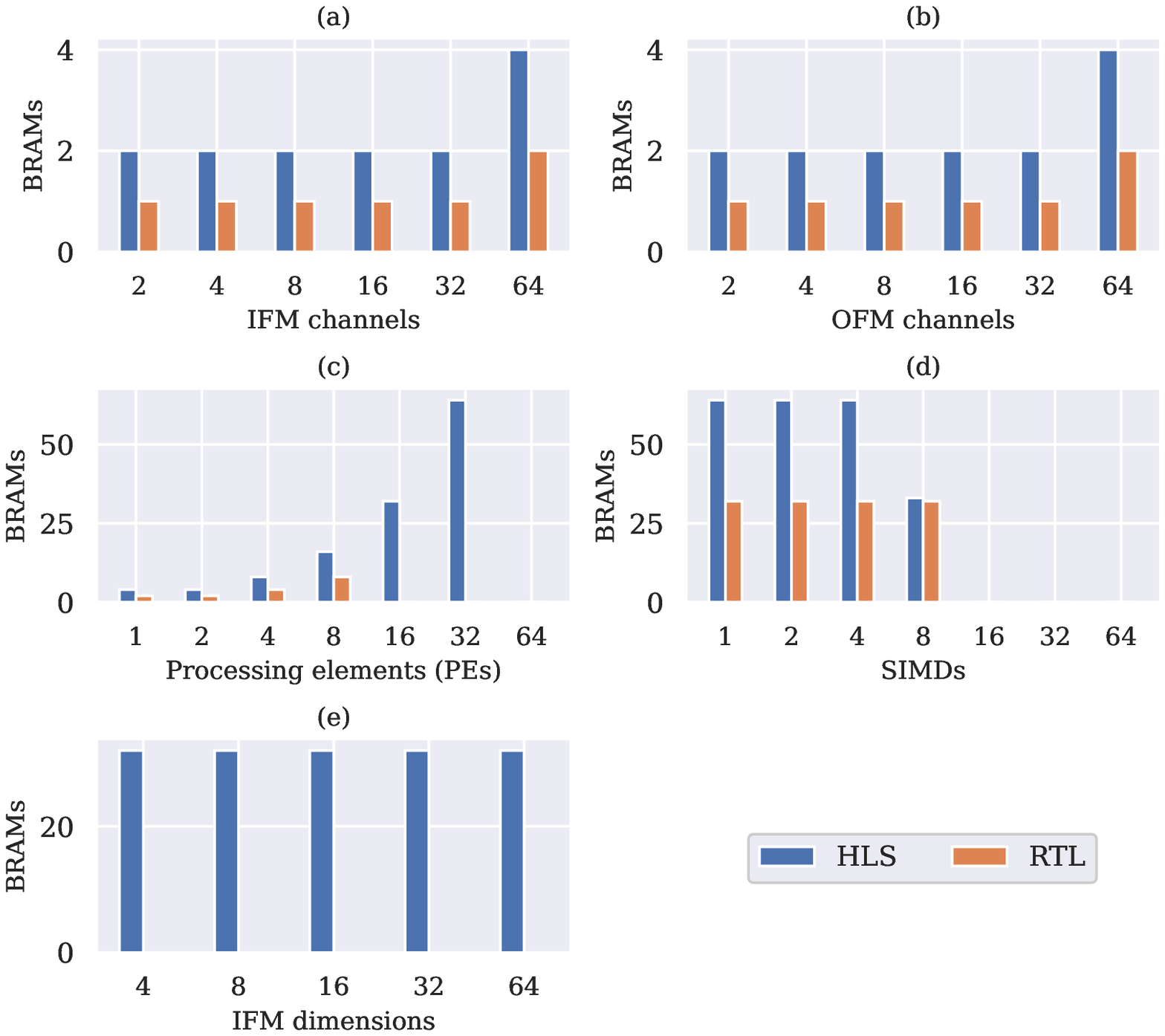}
		\caption{Number of block RAMs (BRAMs) for HLS and RTL when varying different layer and implementation parameters with 1-bit precision for inputs feature map and kernel weights.}
		\label{fig:res_util_bram}
	\end{figure}

	The convergence between HLS and RTL for resource usage is further emphasized by Table~\ref{tab:res_large_design} for the configurations given in Table~\ref{tab:table_large_design} where we increase the number of IFM channels for a larger design with PE=16, and SIMD=16. The effect of an increase in the input buffer depth and base control logic is no longer significant, and HLS and RTL use similar amounts of resources with a trend of HLS eventually even outperforming RTL for larger designs in terms of LUTs. However, HLS always consumes more flip-flops (FFs). This is primarily due to HLS pipelining the generated design aggressively in a pursuit to meet the timing constraints while also attaining an initiation interval (II) of one. The RTL implementation was designed with an II of one to begin with. As it has an explicit cycle-accurate compute schedule by design, its only possible failure mode is the violation of the set clock period constraint.

	\subsubsection{Block RAMs (BRAMs)}
	\label{subsubsec:res_bram}
	
	Block RAMs are dedicated on-chip memory blocks available in FPGAs to facilitate high-speed designs. They come in fixed sizes, e.g., the size of one BRAM tile in the target FPGA of this work is 36~Kb with a total of 4.9~Mb of BRAMs on the device. One 36~Kb BRAM can be partitioned into two 18~Kb memories. As mentioned earlier, the choice of whether to use BRAMs or LUTs for implementing memories was left to the synthesizer. By default settings, HLS is found to use more BRAMs when compared to RTL. In fact, for some cases, RTL does not use any BRAM while HLS ended up using a significant number of them. All of this is illustrated in Fig.~\ref{fig:res_util_bram}(e). The high usage for HLS is due to the unfavourable choice to implement weight memories using BRAMs as they are under utilized. A better alternative will have been to use distributed memory using LUTs.
	
	% \TODO{(Tom) Could I have a look at a synthesized DCP of such an HLS design? I would like to somewhat clear up this speculation. Is it weight memory? Is it buffers? Is it state machine tables?} 
	
	% We can only speculate at the explanation for this high usage. Perhaps, inefficient allocation of BRAMs, preference for BRAMs over LUT RAMs, or perhaps over provisioning when aiming at an II of one while meeting the timing constraints. 
	
	% \TODO{HLS -> unfavorable choice to implement the weight memory using BRAMs, under utilized, good for distributed RAM}
	
	\subsubsection{Resource Utilization: Summary}
	\label{subsubsec:res_util_summary}
	
	Summarizing the resource utilization by HLS and RTL, it is clear that for smaller designs, HLS uses significantly more resources than equivalent RTL designs. In particular, as the number of IFM channels was increased, resulting in an increase in the depth of the input buffer, the HLS resource count increased significantly as it realized a complex multiplexer network for the buffer access. At the same time, RTL resource utilization remained fairly constant. This was consistent across all data types. However, as the core architecture of MVU increases in terms of PEs and SIMDs, there is a convergence between the resource utilization of both. In fact, HLS attains a lower LUT count for a number of designs. With regards to flip-flops though, HLS uses more for all types of designs, which is due to HLS trying to meet the II of one while also meeting the timing constraints. Finally, in regards to block RAMs (BRAMs), HLS mostly uses at least $2\times$ more of these resources compared to RTL.
	
	\subsection{Critical Path Delay}
	\label{subsec:res_delay}
	
	The critical path delay of a circuit determines the maximum clock frequency it can run at. All the designs were properly constrained in terms of clock period as well as signal input and output delays as described in Section~\ref{subsec:experimental}. All the results shown in this section correspond to those shown in Section~\ref{subsec:res_util} for the resource usage.
	
	Table~\ref{tab:res_dly_final} gives the critical path delay, in terms of minimum, maximum and mean delay, of HLS and RTL designs when exploring a range of values of different parameters. When changing the number of IFM and OFM channels, the critical path delay of both RTL and HLS remain consistent, indicated by similar minimum, maximum and mean delay. This is because for HLS, critical path is typically in either in the the adder tree required to add the SIMD outputs or the SIMD elements per PE which remains the same since the number of PEs and SIMDs do not change. For RTL, critical path is in the control logic which also does not change, apart from adjusting to the increase input buffer length with increasing IFM channels. 
	
	Furthermore, RTL designs are consistently faster than HLS. They are around $45\%$ faster for designs with 1-bit precision inputs and binary weights. For the 4-bit precision inputs, the HLS designs become significantly more slower, around $80\%$, as compared to RTL. 
	
	\begin{table}[!t]
		\centering
		\caption{Critical path delay (ns) for HLS- and RTL-based designs.}
		\label{tab:res_dly_final}
		\begin{tabular}{c|c|ccc|ccc}
			\toprule
			\multirow{2}{*}{Paremeter} & \multirow{2}{*}{SIMD type} & \multicolumn{3}{c|}{HLS} & \multicolumn{3}{c}{RTL}\\
			& & Min. & Max. & Mean & Min. & Max. & Mean\\
			\midrule
			\multirow{3}{*}{IFM channels} & XNOR & 2.427 & 2.636 & 2.549 & 1.4 & 1.423 & 1.412\\
			& Bin. weights & 2.445 & 2.641 & 2.567 & 1.4 & 1.424 & 1.413 \\
			& Standard & 7.357 & 7.441 & 7.409 & 1.406 & 1.609 &  1.526 \\\midrule
			\multirow{3}{*}{OFM channels} & XNOR & 2.458 & 2.715 & 2.570 & 1.333 & 1.42 & 1.394\\
			& Bin. weights & 2.476 & 2.651 & 2.560 & 1.421 & 1.424 & 1.422\\
			& Standard & 7.384 & 7.384 & 7.384 & 1.529 & 1.609 & 1.545 \\\midrule
			\multirow{3}{*}{PEs} & XNOR & 2.47 & 2.747 &  2.552 & 1.613 & 2.644 & 1.992\\
			& Bin. weights & 3.842 & 4.009 & 3.917 & 1.833 & 3.327 & 2.598\\
			& Standard & 8.087 & 8.977 & 8.617 & 2.0 & 4.884 & 2.898\\\midrule
			\multirow{3}{*}{SIMDs} & XNOR & 2.244 & 2.936 & 2.667 & 1.437 & 2.644 & 1.859\\
			& Bin. weights & 2.698 & 4.531 & 3.387 & 1.684 & 3.231 & 2.336\\
			& Standard & 7.063 & 9.374 & 8.037 & 1.814 & 3.072 & 2.453\\\bottomrule
		\end{tabular}
	\end{table}
	
	As the number of PEs and SIMDs are increased, the critical path of RTL is also either in the SIMD elements or the adder tree, like that of HLS designs and the critical path delay is directly proportional to the number of PEs and SIMDs. Thus, the delay increases for both HLS and RTL indicated by increase in the  maximum and mean critical path delay as these two design parameters are increased. In all cases, RTL designs are $45\%$ -- $75\%$ faster than HLS designs.
	
	\subsubsection{Critical Path Delay: Summary}
	\label{subsubsec:res_dly_summary}
	
	To summarize, the RTL designs are faster than HLS designs for all types of networks. The critical path is proportional to the number of PEs and SIMDs. Thus, change in number of IFM and OFM channels do not affect the critical path delay. Apart from a change in the input buffer depth, the overall design complexity remains the same. However, increase in design complexity by increasing either PE or SIMD or both leads to increase in the critical path delay.
	
	% networks with 1-bit precision inputs and binary weights (with non-binary activations) while RTL designs are faster for standard, non-binary inputs. In contrast to HLS, the RTL timing is much more predictable and, in fact, largely constant as different network and design parameters are varied.
	
	\subsection{Synthesis Times}
	\label{subsec:syntimes}
	
	%%% Results for Tool execution time
	The overall design time needed for an RTL implementation is significantly higher than that needed for HLS. However, it is important to analyze how long it takes to synthesize the two design approaches and to see whether the benefits achieved with a shorter design time for HLS can be carried over to actual synthesis time.
	
\begin{figure}[!t]
		\centering
		\includegraphics[scale=0.45]{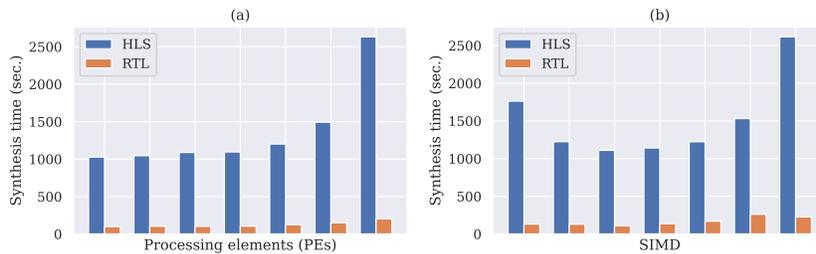}
		\caption{Total synthesis time for HLS and RTL when varying the number of PEs and SIMDs/PE.}
		\label{fig:exec_time}
	\end{figure}

	For the results shown earlier, tool execution time was measured for both HLS and RTL when modifying various network and design parameters. Only compilation and synthesis times were considered and results are shown in Fig.~\ref{fig:exec_time}
	
	It can be clearly seen that the synthesis time of HLS is at least $10\times$ more than RTL with a superlinear growth for larger designs. For other network parameters, like the number IFM and OFM channels, the dimension of the IFM and filter kernel, HLS also uses significantly more time to synthesize, though they are not shown here. However, the HLS synthesis time does not grow significantly as we modify other parameters, mainly because they either only impact the depth of the input buffer or have no impact at all on the overall design complexity of the MVU. In addition to this, HLS synthesis time was a limiting factor towards synthesizing larger designs than those presented in this work.
	
	\subsection{Full Example: Network Intrusion Detection (NID)}
	\label{subsec:res_app}
	
	High-performance packet processing systems, as for network intrusion detection (NID), are typically implemented on FPGAs because of the inherent parallelism of the FPGA architecture and the streaming nature of the accelerators which can deliver far lower latency. Such systems enable increased network security and may be implemented using deep neural networks (DNNs) \cite{Umuroglu_2020C}. The DNN used for this NID is a multi-layer perceptron (MLP) network of four layers, the parameters of which are given in Table~\ref{tab:table_app}. Depending on the throughput requirements, different configurations of PEs and SIMDs/PE can be used and the ones used in this work are also given in Table~\ref{tab:table_app}.
	
	\begin{table}[!t]
		\centering
		\caption{Configuration parameters for multi-layer perceptron (MLP) network of 4 layers for intrusion detection.}
		\label{tab:table_app}
		\begin{tabular}{llcccc}
			% \cline{3-8}
			% & & \multicolumn{6}{c}{Variable parameters} \\
			% \cline{3-8}
			% & & IFM channels & IFM dimensions & OFM channels & Kernel dimensions & PE & SIMD\\\cline{3-8}
			% \parbox[t]{2mm}{\multirow{5}{*}{\rotatebox[origin=c]{90}{Fixed parameters}}} &
			\toprule
			Layer & 0 & 1 & 2 & 3\\
			\midrule
			IFM channels & 600 & 64 & 64 & 64 \\
			IFM dimensions & 1 & 1 & 1 & 1 \\
			OFM channels & 64 & 64 & 64 & 1 \\
			Kernel dimensions & 1 & 1 & 1 & 1\\
			Weight precision & 2 & 2 & 2 & 2 \\
			Input precision & 2 & 2 & 2 & 2 \\
			PE & 64 & 16 & 16 & 1 \\
			SIMD & 50 & 32 & 32 & 8\\
			\bottomrule
		\end{tabular}
	\end{table}
	
	The results of the synthesis of the four layers with the given network and hardware parameters of Table~\ref{tab:table_app} are presented in Table~\ref{tab:res_app}. We see the same behavior as earlier: RTL produces smaller circuits than HLS for smaller designs but not for larger ones. RTL designs, however, have an improved critical path delay and a significant reduction in synthesis time. However, the synthesis time needs to be seen in the context of the larger design effort to write RTL to describe an architecture. The execution cycles for both RTL and HLS are fairly similar and both achieve an initiation interval (II) of one. 
	
	Thus, the findings of earlier experiments are re-affirmed by evaluating a real-world use case of network intrusion detection using an MLP network.

	\begin{table}[!t]
		\centering
		\caption{NID synthesis results for HLS and RTL.}
		\label{tab:res_app}
		\begin{tabular}{l|cc|cc|cc|cc|cc|cc}
			\toprule
			\multirow{2}{*}{Layer} & \multicolumn{2}{c|}{LUTs} & \multicolumn{2}{c|}{FFs} & \multicolumn{2}{c|}{BRAM} & \multicolumn{2}{c|}{Delay (ns)} & \multicolumn{2}{c|}{Synth. time} & \multicolumn{2}{c}{Exec. cycles}\\
			& HLS & RTL & HLS & RTL & HLS & RTL & HLS & RTL & HLS & RTL & HLS & RTL\\
			\midrule
			Layer \#0 &  30744 &  43894 & 21159 & 12965 & 0 & 0 & 7.081 & 5.292 & 38'45'' & 5'21'' & 17 & 17\\
			Layer \#1/2 &  4653 & 5454 & 3276 & 4970 &  0 & 0 & 7.453 & 4.959 & 17'48'' & 3'59'' & 13 & 13\\
			Layer \#3 &  248 & 133 & 364 & 158 & 0 & 0& 7.132 & 4.959 & 16'28'' & 1'43'' & 12 & 13\\
			\bottomrule
		\end{tabular}
	\end{table}

	\section{Conclusion}
	\label{sec:conc}
	
	High-level synthesis (HLS) has become a popular alternative to the design of digital systems on the register-transfer level (RTL) using hardware description languages (HDL). HLS provides flexibility to designers, especially to software engineers, and allows them to describe their designs in a high-level language like C++. It reduces the design time and requires less specialized training as compared to RTL design entry. Over the years, HLS has improved significantly, and the performance gap between HLS and RTL has reduced.
	
	HLS has been used successfully to describe designs of various applications like high-performance computing and neural networks. The FINN framework is one such example. It generates DNN accelerators targeting FPGAs by producing synthesizable C++ HLS code. These accelerators are constructed as streaming dataflow (DF) architectures and leverage reduced-precision data types. The main compute units in FINN accelerators are instances of the matrix-vector unit (MVU). They multiply input feature vectors with the kernel weight matrices of a neural network layer. This work presented an alternative RTL description of this MVU and evaluated the attained design trade-offs in comparison to the baseline HLS implementation.
	
	The RTL description is tailored to the requirements of the MVU and is designed to achieve an initiation interval (II) of one. By implementing an AXI-stream input/output interface, it is a drop-in replacement for the generated HLS cores. The design comparison, which was conducted for a Zynq-7000 FPGA target platform, analyzed the resource utilization, the critical path delay and the needed synthesis time. The resource utilization was further detailed into look-up table (LUT), flip-flop (FF) and block RAM (BRAM) usage.
	
	The analysis was carried out by sweeping through a number of network and design parameters, namely, the number of input feature map (IFM) and output feature map (OFM) channels, the dimensions of the IFM and the kernel as well as the numbers of processing elements (PEs) and SIMD slices per PE. It was shown that HLS requires significantly more FPGA resources as compared to RTL for smaller designs due to a large base implementation to cater for I/O protocols, buffers and overall architecture. Increasing the design complexity, i.e. the number of PEs and SIMDs, lets the resource utilization figures of HLS and RTL designs, particularly the LUT usage, converge. However, HLS consistently consumes orders of magnitude more flip-flops and at least $2\times$ more BRAMs than the RTL design. This is caused by the HLS synthesis aggressively pipelining the generated RTL code as a proactive measure for reducing the risk of later timing violations. Finally, it was demonstrated that the resource consumption by the HLS design is particularly sensitive to the number of IFM channels. The increase of this parameter mandates larger input buffers, which the HLS design accesses through an unfavorably growing multiplexer structure.
	
	The critical path of the MVU is in the control logic for RTL and in the SIMD elements or the adder tree for HLS for small number of PEs and SIMDs. As the design becomes larger, the critical path of the RTL designs is also in the SIMD elements or the adder tree and the delay is directly proportional to the number of PEs and SIMDs. This results in similar delay through the circuit as the number of IFM and OFM channels are changed while keeping the number of PEs and SIMDs constant. As the design becomes larger with an increase in the number of PEs and/or SIMDs, the critical path delay increases for both RTL and HLS designs. In all cases, however, The RTL produces faster designs, between $45\%$ -- $80\%$,  for all types of networks as compared to HLS.
	
	HLS also suffers from significantly longer synthesis times compared to RTL. This was a limiting factor on the size of designs that can be synthesized. It took at least $10\times$ more time to synthesize an HLS design as compared to RTL. This has the potential to undo any gains achieved in the reduction of the original design time when the design space exploration of debugging demand additional complete design cycles.
	
	To conclude, the RTL abstraction is an attractive alternative for code-generated hardware designs of frameworks such as FINN and hls4ml, given the gains achieved in synthesis time together with some potential resource benefits depending on the design size.

	\section*{Acknowledgements}
	This material is based upon work supported, in part, by Science Foundation 
	Ireland under Grant No. 13/RC/2094\_P2 and, in part, by the 
	European Union's Horizon 2020 research and innovation programme under the Marie 
	Sk\l odowska-Curie grant agreement and Grant No. 754489. Any opinions, findings, 
	and conclusions or recommendations expressed in this material are those	of the author and do not 
	necessarily reflect the views of the Science Foundation Ireland and European Union's Horizon 2020 
	programme.
	
	
\end{document}